# CancerLinker: Explorations of Cancer Study Network

Vinh Nguyen, Md Yasin Kabir, and Tommy Dang

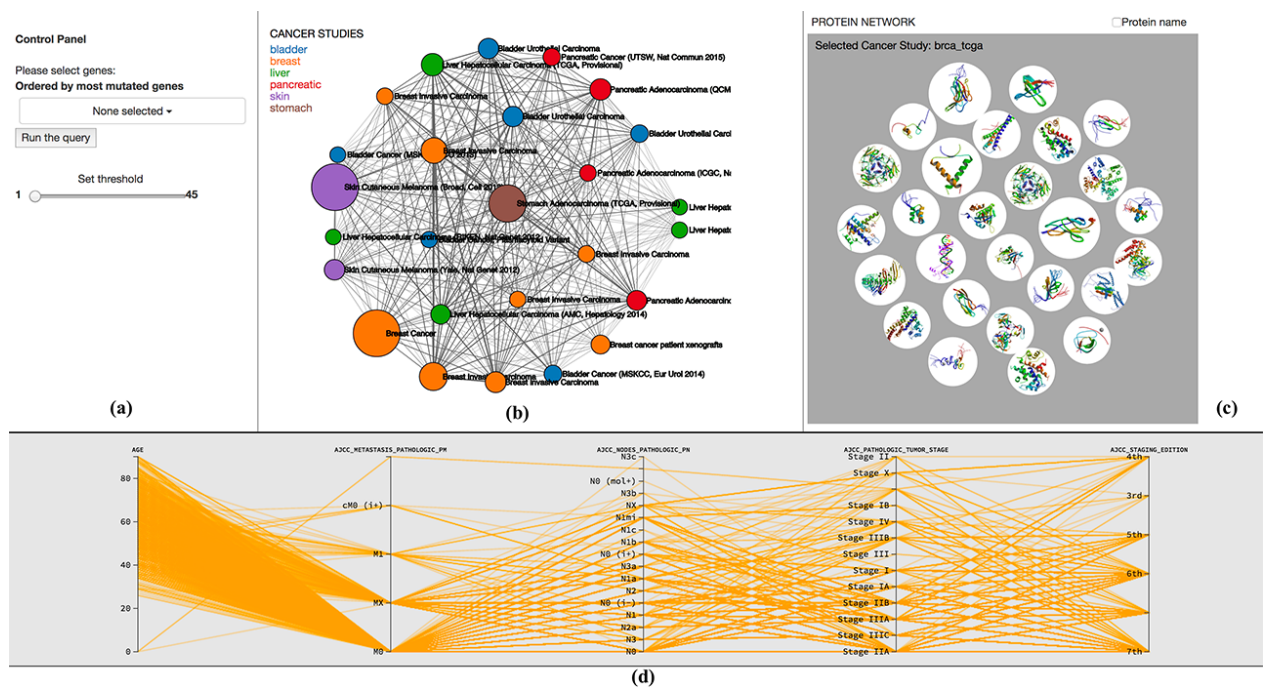

Figure 1. Overview of CancerLinker; a) Control panel of CancerLinker, b) The merged network of twenty-six cancer studies, c) Bubble chart of proteins within a cancer study. d) The parallel coordinate highlighting patterns of patient profiles

**Abstract**—Interactive visualization tools are highly desirable to biologist and cancer researchers to explore the complex structures, detect patterns and find out the relationships among bio-molecules responsible for a cancer type. A pathway contains various bio-molecules in different layers of the cell which is responsible for specific cancer type. Researchers are highly interested in understanding the relationships among the proteins of different pathways and furthermore want to know how those proteins are interacting in different pathways for various cancer types. Biologists find it useful to merge the data of different cancer studies in a single network and see the relationships among the different proteins which can help them detect the common proteins in cancer studies and hence reveal the pattern of interactions of those proteins. We introduce the CancerLinker, a visual analytic tool that helps researchers explore cancer study interaction network. Twenty-six cancer studies are merged to explore pathway data and bio-molecules relationships that can provide the answers to some significant questions which are helpful in cancer research. The CancerLinker also helps biologists explore the critical mutated proteins in multiple cancer studies. A bubble graph is constructed to visualize common protein based on its frequency and biological assemblies. Parallel coordinates highlight patterns of patient profiles (obtained from cBioportal by WebAPI services) on different attributes for a specified cancer study.

✦

## 1 INTRODUCTION

Biologists spend years trying to understand the complex relationships of various cell functions and inter-networks between enormous numbers of components of the cells [14]. It is a very arduous task to explore the cell components and reveal the patterns, anomalies, relations and other signification information just from data which are mostly texts and statistics. However, a friendly and efficient visualization of biological data might help to understand that arduous task in a more productive way.

Visualization is the process of representing data visually and is a signification task in network analysis, especially in the biological system where scientists work with the complex structures and relationships of the bio-molecules. Hence, there are plethora visualization tools available in recent years to utilize the power of visualization to help cancer researchers better understanding of the complex systems such as pathways, protein networks, etc. Despite such available tools and techniques, understanding protein-protein interactions is still ongoing challenges for biologists and researchers for years, even with a small question of how to better understand complicated many-to-many relationships between cancers and mutated genes. In addition, existing tools are often limited to some kinds of data. Therefore, having a graphical tool that helps biologists analyse and gain insight interactions of the protein-protein network is crucial.

A pathway represents the complex relationship of biochemical interaction among the molecules inside a cell [13]. Each cell contains an enormous number of bio-molecules responsible for every interaction and every incident happening in that cell. The division of a cell, the lifetime of a cell, and how cell functions depend on the bio-

- *Vinh Nguyen. E-mail: vinh.nguyen@ttu.edu*
- *Md Yasin Kabir. E-mail: yasin.kabir@ttu.edu.*
- *Tommy Dang. E-mail: tommy.dang@ttu.edu.*

molecules contained by that cell. Those bio-molecules, might locate in different layers of a cell, perform their activities in a linear and cyclic order. If by any signals or another way the first component of a cell becomes active then it triggers the other molecules to become active and carries the signal till the end of the process. Any abnormality of this process can lead a cell become a cancer cell. Hence, understanding the pathways and the relationship between molecules is crucial in a cancer research.

In this paper, we propose *CancerLinker*, a visual analytic tool, to explore the relationships among genes (which are essentially proteins or part of proteins) in different cancer studies for providing an interactive way to help researchers understand the complicated many-to-many relationships between cancers and mutated genes.

## 2 RELATED WORK

A pathway incorporated with hundreds of bio-molecules in a cell. Those bio-molecules perform and control the different activities and interaction in that cell. If the first molecule in the pathway receives a signal from outside of the cell, it starts to interact with others and activates other molecules respectively till the last molecule is activated and the cell function is completed. The order of interaction and activation is modelled as feedback loops [8] where the output from one molecule is work as the input of another molecule. Unusual signal or abnormal activity in the pathway can lead to cancer. That's why it is crucial to study and learn the pathways of the different cells. The relationship between molecules in a pathway is dynamic and intricate. Understanding those relationships and enabling a way to view higher order pattern is a remarkable challenge in modern bioinformatics research [17]. Many visualization techniques and tools have been proposed to address these challenges. PathwayMatrix [9] provided a comprehensive overview of the binary relation between proteins in an adjacency matrix. It can sort proteins based on their similarities and offer a clustering of identical proteins which is arduous to detect using traditional node-link diagrams like PCViz [6] or VisANT [12]. Although matrix representation is offering some advantages over traditional techniques, however, a drawback is that it does not show the path between nodes.

*Cystoscape* [20] is an open source, standalone Java application. It allows to merge and visualize data quickly using a friendly interface. It helps to configure, visualize and import data very efficiently with different plugins developed by the various communities, including Cerebral [2] and RenoDoI [22]. The disadvantage of Cystoscape in cancer study is not allowing users to visualize the data by specific parameters. For example, differentiating cancer studies by colours is not offered in this application.

*Medusa* [11] is another open source system written in Java programming language and available as an applet. Although it is highly interactive and provides a lot of algorithms to perform on a network, it does not allow users to view different types of visualizations side by side and perform some operations on one visualization based on the interaction of another.

*Gephi* [3] is an interactive visualization tool for all kind of complex networks; it is a Java application that supports multiple data files. However, it does not help viewing two different networks or visualization together with interaction.

*Networkx* [18] is a Python language software package for the creation, manipulation, and study of the structure, dynamics, and functions of complex networks. This application requires high-level skills to use. It uses command line interface, and it lacks interactive visualization.

In this paper, we introduce *CancerLinker*, a novel exploration tool, that allows users to uncover the correlations among cancer studies and genes. This visual system provides users with handful options to explore cancer studies, detects relationships among those studies, views the assembly structure of proteins, highlights typical patterns of patients' clinical data, and finds out the most active genes in different cancer studies.

## 3 CANCERLINKER VISUALIZATION

We worked closely with a molecular biologist to find out the requirements of the application. We pointed out some of the most important research questions such as: Is there any relationship between any pairs of proteins? What is the strength of these relationships? What is the structure of the protein being studied? What is the pattern of genomic profile data for one or more genes? Or which protein plays an important role across cancer studies? Based on his research questions, we broke down the system requirements into four visualization tasks with the following objectives:

- Visualization task **T1**: To uncover the correlations among cancer studies as well as cancer studies and genes.
- Visualization task **T2**: To explore the role of proteins across cancer studies.
- Visualization task **T3**: To visualize common protein based on its frequency and biological assemblies.
- Visualization task **T4**: To highlight the patterns of patient profiles (obtained from *cBioportal* by WebAPI services) on different attributes for a specified cancer study.

We implemented and tested these objectives on 26 cancer studies which contained 192 proteins.

### 3.1 Processing input data sets

Data for CancerLinker was retrieved from three different sources. The first source was from *cBioPortal* [10] portal where 26 cancer studies were downloaded separately and merged into a single file in CSV format (Comma Separated Value). This merged file includes proteins name, frequency, cancer studies and study type. *cBioPortal* also provides a web API service for retrieving data, we used this service to pull patient clinical data based on cancer study. The output of the API query is in TSV format (Tab Separated Value). The second data source comes from Protein Data Bank (http://www.rcsb.org/) [4] where we got the assembly of a protein based on its name. The final data source is KEGGPATHWAY database [1] where we retrieve connections among proteins.

### 3.2 CancerLinker Overview

In CancerLinker, the application displays 26 different cancer studies and contains five main components: the control panel, the cancer studies' network, the biological assemblies view, the parallel coordinate view, and the two separate communities of the protein network. Figure 1 represents the basic overview of our visualization tool which is implemented using D3js [5]. The control panel is placed on the top left (a), merged network of 26 cancer studies is in the middle (b), the bubble chart of the proteins is shown on the top right(c), (d) represents the parallel coordinate view, and communities are shown on the different page due to the limitation of the screen.

**The Control Panel**

In order to find out which cancer studies contain the desired proteins being investigated, the application provides a search box that allows users to search and select the desired proteins. Proteins are ordered by their frequency and are all selected by default. When biologists want to narrow down the number of proteins found in most cancer studies, a filter slider is provided. The number on the left bar indicates the minimum number of proteins found between cancer studies while the number on the right bar shows the maximum proteins. It can be seen that there is at least one protein found in all cancer studies and 45 is the biggest number of proteins found between cancer studies. The result of these options provides an important information for biologists since they know which proteins are mostly active in different cancer studies.

**The cancer studies network**

The cancer studies network depicted in Figure 2(a) presents the merged network of 26 cancer studies. Each cancer study is represented by a circle. The size of the circle indicates the number of proteins found.

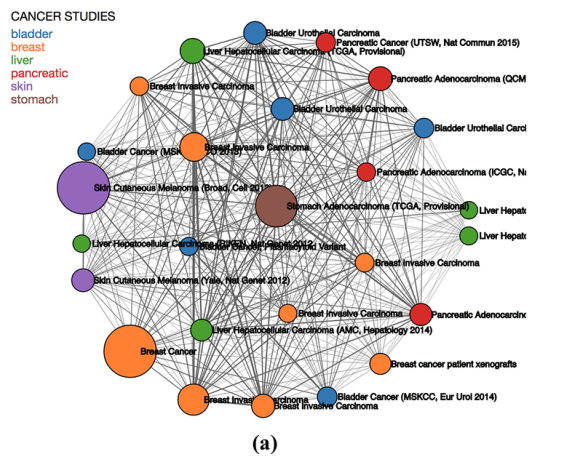

(a)

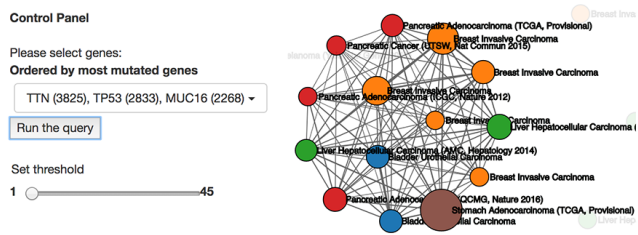

(b)

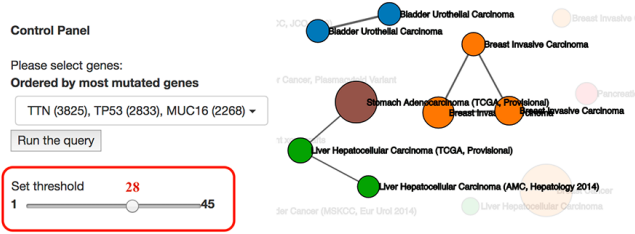

(c)

Figure 3. The cancer studies network; (a) The relationships among 26 cancer studies are representing in a merged network. (b) The filtered cancer studies network based on the selection of three genes. (c) The cancer studies consist at least 28 common mutated genes

The bigger size, the more proteins. Cancer studies are also encoded by colours, each colour represents one type of cancer, including bladder cancer, breast cancer, liver cancer, pancreatic cancer, skin cancer, and stomach cancer. Cancer studies are connected to each other by a line or link. The thickness of the link between any pair of studies indicates the number of proteins found between these studies. The thinker the link, the more protein found in both cancers. Figure 2(b) shows the filtered network based on the selected genes from the control panel on the left: TTN (3825), TP53(2833), and above three genes in common is visible while irrelevant cancer studies are faded out.

In the Figure 2(c), we provide an example by setting the threshold value as of 28 using the slider (inside the red box). CancerLinker highlights only the relations of the cancer studies which have at least 28 mutated genes in common.

**The biological assemblies view**

The biological assemblies view highlights mutated proteins in a specific cancer study (visualization task **T3**). Figure 3 shows an example of selecting the *Cancer Genome Atlas Breast Invasive Carcinoma* [21] from the cancer network in Figure 2(b). In particular

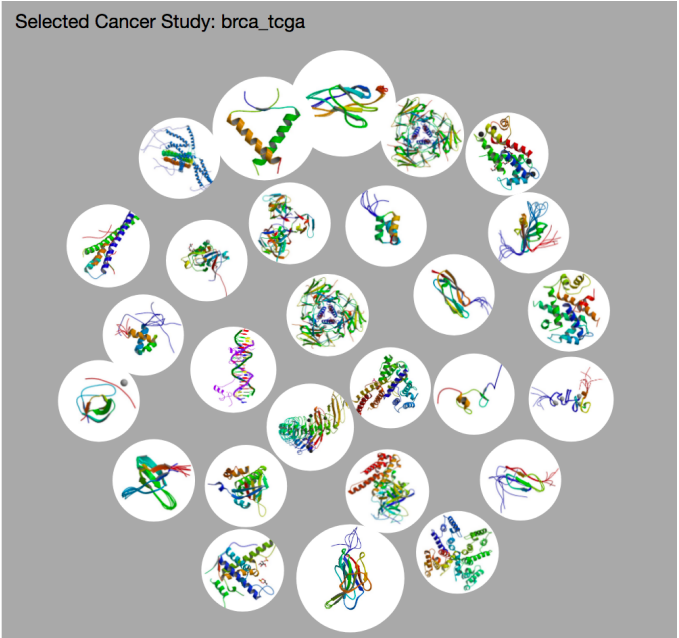

(a)

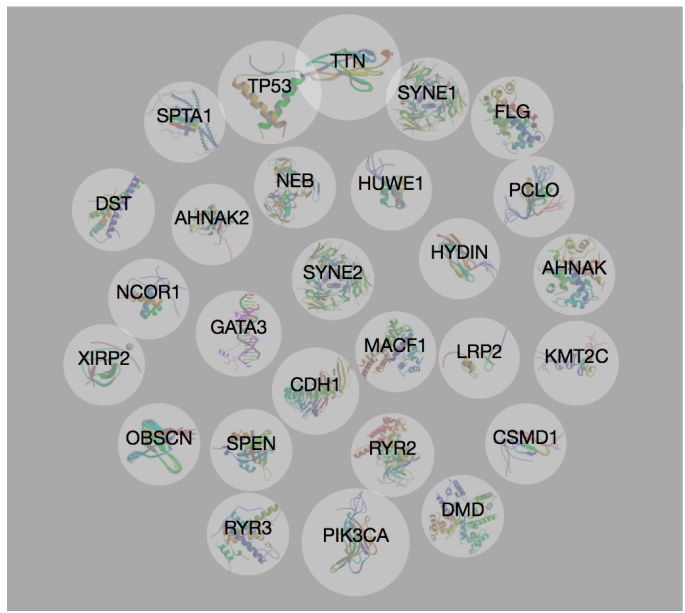

Figure 2. The biological assemblies view of the Cancer *Genome Atlas Breast Invasive Carcinoma* [21]: (a) proteins structures (b) proteins names. Larger bubbles indicate proteins with higher mutated rates.

Figure 3, as users make a selection, the bubble graph in Figure 3(a) is populated with proteins images fetched on the fly from protein data bank [4].

A check-box is provided to enable quickly switching to the respective proteins name as depicted in Figure 3(b). Notice that image sizes are computed based on protein mutation frequency within the selected study. Similarly, when users select a link in the cancer network, the bubble graph displays only proteins which are frequently mutated in both cancers.

**The parallel coordinate views**

It is paramount to know the patient clinical information from which a statistical idea can be acquired. To meet the needs for this requirement, we constructed the parallel coordinate view which is depicted in Figure 4 due to the nature of high dimensional data. It helps users explore the patterns of patient clinical data on a particular cancer study (visualization task **T4**). Information is retrieved from

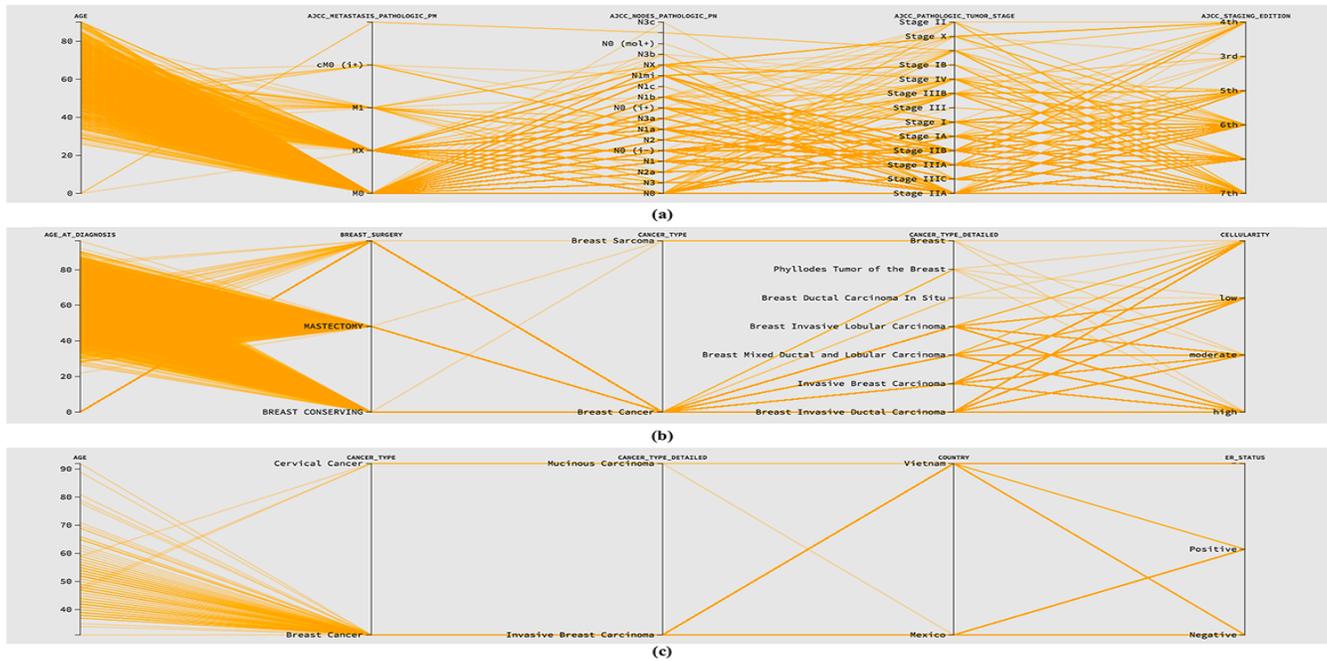

Figure 4. The parallel coordinate view highlights the patient profiles of: (a) Breast invasive carcinoma study shows the various stages of cancer of the patients, (b) General Breast cancer shows the cancer type, cancer type details, cellularity, and (c) Breast Invasive Carcinoma study consists of 103patients of two countries: Vietnam and Mexico.

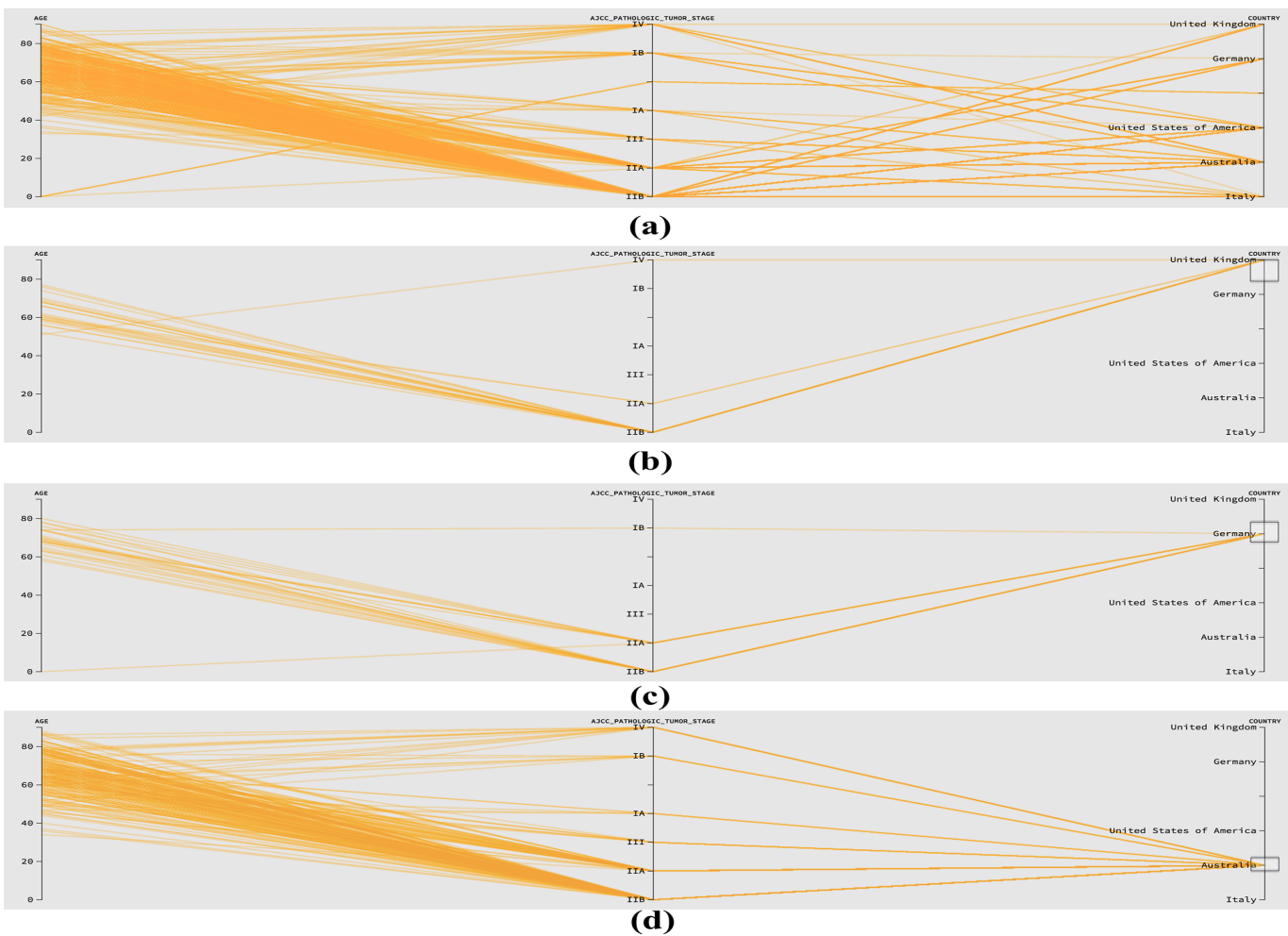

Figure 5. Filtering on parallel coordinates: (a) Pancreatic adenocarcinoma cancer study (b) The selection of United Kingdom, (c) Germany, and (d)Australia.

cBioportal by a web API service and the number of attributes varies in each cancer study. Typically, the output from this query includes information of patients such as age, cancer type, stage of cancer, location, gender and some more details. The parallel coordinate supports brushing which enables users to select any point or multiple point from a specific bar to see the distribution of patient data.

Figure 4(a) represents the parallel coordinate view for *Breast invasive carcinoma study* which was performed on 1098 patients (1098 polylines). The view represents the ages, pathological information tumor stage, and staging edition. The view also enables to see the relationship between ages and other information of this cancer patients. Figure 4(b), shows the information from the study of General breast cancers which consists of age, sub-cancer types, and cellularity information of 2509 patients. The parallel coordinate view of *Breast Invasive Carcinoma study* in Figure 4(c) represents the cancer types, sub-cancer types, countries, and ER-status of the patients. This view can help to compare the sub-cancer types and also related the location information where a particular cancer type can be more severed.

CancerLinker supports a wide range of interactions on parallel coordinate view. The Figure 5 shows the usefulness of selection tool in parallel coordinate view. The figure represents the information of the patients of *Pancreatic adenocarcinoma cancer* study of different age and countries. The top image Figure 5(a) shows the parallel coordinate view of *Pancreatic adenocarcinoma cancer* study. The second image Figure 5(b) provides an idea of the number of patients who are from the United Kingdom and their ages. The third shows the information for the Germany. The number of patients from both countries was 23 and 26 respectively. However, an interesting finding from this parallel view is the age range of the patients in both countries. In Germany, the patients who have this particular cancer are ages from around 57-80 years. On the other hand, In United Kingdom, the years flows between 50-77. If we look at the Australia, we can see that most of the patients of *Pancreatic adenocarcinoma cancer* are from Australia (exactly 216 out of 383) and the ages of the patients are distributed between 30-90. All of those views denote that *Pancreatic adenocarcinoma cancer* mostly found on adult above 30 years old.

**Community detection for protein network**

The requirement of Visualization task **T2** leads us to the idea of constructing community networks to answer to question which proteins are highly active or interacts with many different proteins or a group of proteins tends to interact with one another more than others.

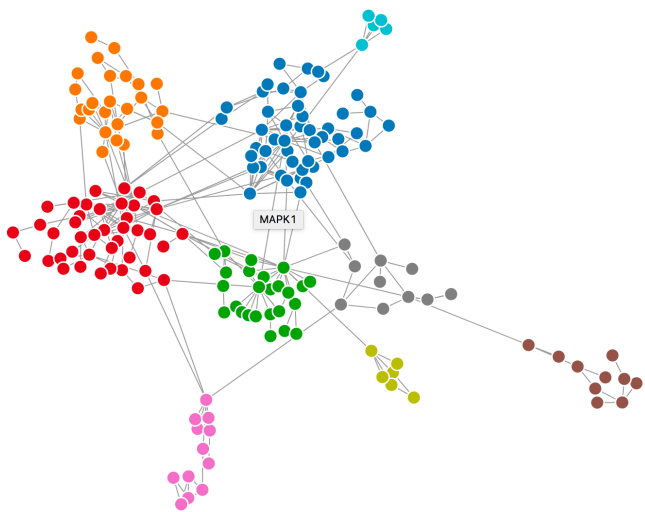

Figure 6. The 9 communities of the protein interaction network using edge betweenness algorithm [16]. Colors encode proteins in the same cluster.

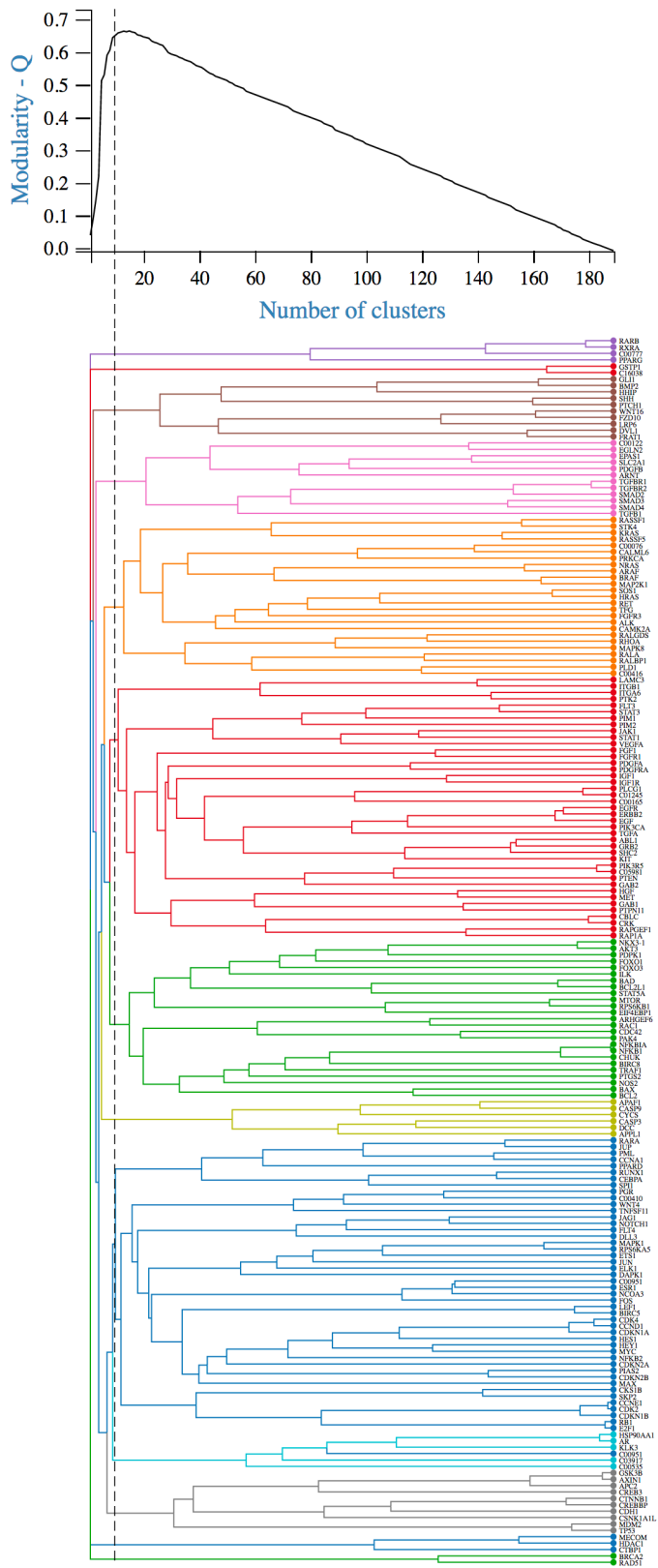

Figure 7. Dendrogram constructed by using the Newman and Girvan edge betweenness [16] algorithm for the network in Figure 6. The dashed slider is used to control the number of cluster.

Figure 6 depicts clusters of 192 proteins from 26 cancer studies. We further applying forces to group the nodes of the same community's closer to provide a more coherent view. In this example, we use edge betweenness algorithm [16] of Newman and Girvan. This algorithm computes the pair dependencies value for all the edges using shortest paths in the first step and removes the edge with most betweenness centrality in next step. After that, it again reconstructs the network and recalculate the betweenness centrality and continue until there is no edge. Using those steps, the algorithm creates a hierarchical structure, named *dendrogram*. After that using modularity [15], we find out the best cut-off value for detecting community structure within the protein interaction network.

Figure 7 represents the dendrogram constructed by using the Newman and Girvan edge betweenness [16] algorithm for the network in Figure 6. We can drag the dashed bar to adjust the number of communities. The color of the dendrogram is consistent to the color of communities in Figure 6. In this example, we have selected the number of cluster is 9, which is also the best modularity *Q* for this network (as indicated in the modularity graph on the top). As users move the dashed slider (to select a different number of cluster), the main network. The merged protein network can be shown at different viewpoints as depicted in Figure 8. Using Newman and Girvan algorithm, edge betweenness centrality value for each node is calculated, the size of each circle represents its betweenness centrality value. The bigger size, the bigger value and there more important it is. A viewer is also able to see the connection of protein with other

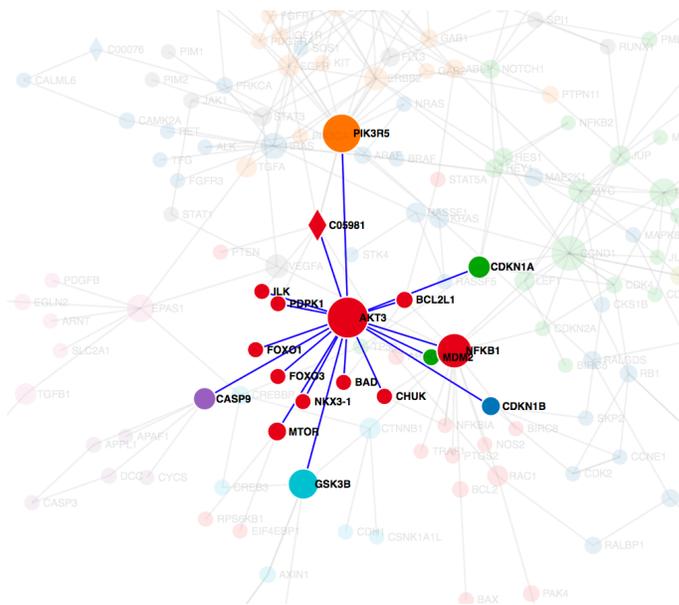

Figure 9. The direct connections of a protein with other proteins and compounds. Node size represents the importance of a protein based on the direct relation to other proteins and compounds. We calculate the importance of a node using edge betweenness centrality [16].

Colour, shape and size of the graph and network were set based on user's preferences.

With the support of visualization tool, biologists can quickly get insights of the studies. Results showed that that Breast Cancer accounts for the biggest proportion of protein found in all cancer studies. AKT3, PIK3R5, CCND1 and NFKB1 are the most important proteins in the network since they interact with many other proteins and hold the position in the network as a bridge.

Drilling down into the Parallel coordinates, users found out that in Germany, the patients who have this particular cancer are ages from around 57-80 years. While in United Kingdom, these years flows between 50-77. If users look at the Australia, they can see that most of the patients of *Pancreatic adenocarcinoma cancer* are from Australia (exactly 216 out of 383) and the ages of the patients are distributed between 30-90. All of those views denote that *Pancreatic adenocarcinoma cancer* mostly found on adult above 30 years old.

Summative evaluation was conducted after the project was fished, reports from biologist suggested that the protein network should be improved in such a way that users can quickly refer back to the cancer study of the protein without having to click on it. We proposed a solution for this suggestion by replacing circles with pie charts. Each slice of the pie will be represented by a cancer study and colour-encoded corresponding to its studies.

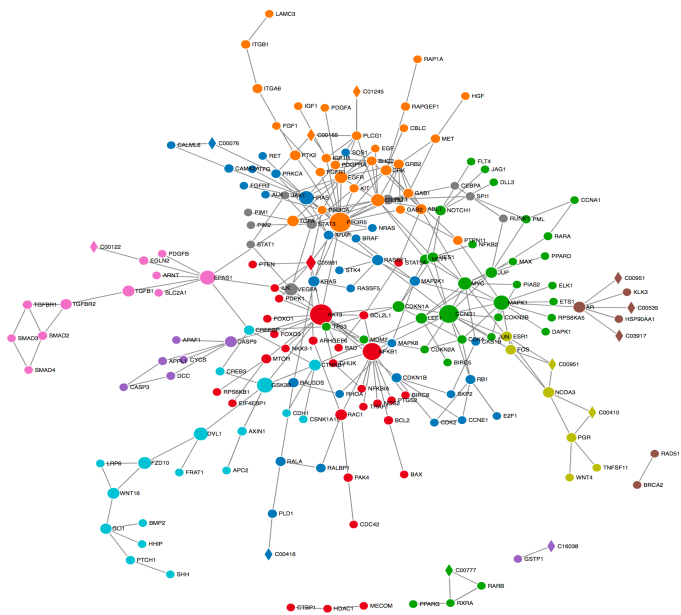

Figure 8. The communities of the proteins network of the cancer. The circular nodes represent the proteins, and the diamond shape nodes represent the compound in a cancer pathway.

proteins and compounds by clicking on any proteins in the community structure as depicted in Figure 9.

## 4 IMPLEMENTATION

CancerLinker is implemented in JavaScript with D3 library [5]. The code is open source and will be available on our *Github* link with demo video and project documentation at:
*https://github.com/iDataVisualizationLab/CancerLinker*.

## 5 RESULTS AND EVALUATION

In our application, we conducted both formative evaluation and summative evaluation methods to evaluate the effectiveness of the application. By working closely with a biologist at the beginning of the stage, all research questions were converted into visualization tasks. The application was built based on these provided tasks.

## 6 CONCLUSIONS AND FUTURE WORK

Visualizing complex biological network like cancer studies is an ongoing challenge and an exciting research field which can help significantly to detect patterns and relationships. Visualization tools can help to track the relations and interactions between different molecules in different stages and times. The aim of CancerLinker is to assist a researcher in finding patterns between cancer studies and genes. It also helps to explore the patients' data which can help to get a demographic idea about a particular cancer type.

As depicted in this paper, experimental data contain many variables (multidimensional data). Looking into the relationships of different variables in experimental data may lead to interesting discoveries. However, inspecting individual dimension as well as the correlations between dimension in parallel coordinates is a time-consuming process. In the future, we plan to apply visual features [19, 23] to highlight interesting scatter-plots (and variables), for examples, scatter-plots with clusters [7] representing different classes of cancer patients. Using visual features, we should be able to extract important variables (and genes/proteins) corresponding to the separation of different groups of cancer patients. Highlighting these genes/proteins (using different colour encoding) on the biological maps can be significant in analysing the causality inherent in biological networks and thus very valuable in drug design [14].